\documentclass[aps,prd,nofootinbib,floatfix,superscriptaddress,secnumarabic]{revtex4}
\usepackage{amsmath}
\usepackage{graphicx}
\usepackage{xcolor}
\usepackage[normalem]{ulem}
\newcommand{\MSbar}{{\overline{\rm MS}}}
\newcommand{\pa}{\partial}
\newcommand{\gtilde}{\frac{g^2}{16 \, \pi^2}\; }
\newcommand{\qslash}{{\not{\hspace{-0.05cm}q}}}

\newcommand{\la}{\lambda}

\newcommand{\be}{\begin{equation}}
\newcommand{\ee}{\end{equation}}
\newcommand{\bea}{\begin{eqnarray}}
\newcommand{\eea}{\end{eqnarray}}

\begin{document}

\title{Supersymmetric QCD on the lattice: Fine-tuning and counterterms for the quartic couplings}

\author{M.~Costa}
\email{marios.costa@cut.ac.cy}
\affiliation{Department of Mechanical Engineering and Material Science and Engineering, Cyprus University of Technology, Limassol, CY-3036, Cyprus}
\affiliation{Rinnoco Ltd, Limassol, CY-3047, Cyprus}
\affiliation{Department of Physics, University of Cyprus, Nicosia, CY-1678, Cyprus}

\author{H.~Herodotou}
\email{herodotos.herodotou@ucy.ac.cy}
\affiliation{Department of Physics, University of Cyprus, Nicosia, CY-1678, Cyprus}

\author{H.~Panagopoulos}
\email{haris@ucy.ac.cy}
\affiliation{Department of Physics, University of Cyprus, Nicosia, CY-1678, Cyprus}

\begin{abstract}
In this work we calculate the renormalization of counterterms which arise in the lattice action of $\mathcal{N} = 1$ Supersymmetric QCD (SQCD). In particular, the fine-tunings for quartic couplings are studied in detail through both continuum and lattice perturbation theory at one-loop level. For the lattice version of SQCD we make use of the Wilson gauge action for gluon fields and the Wilson fermion action for fermion fields (quarks, gluinos); for squark fields we use na\"ive discretization. On the lattice, different components of squark fields mix among themselves and a total of ten quartic terms arise at the quantum level.  Consequently, the  renormalization conditions must take into account these effects in order to appropriately fine-tune all quartic couplings. All our results for Green's functions and renormalization factors exhibit an explicit analytic dependence on the number of colors, $N_c$, the number of flavors, $N_f$, and the gauge parameter, $\alpha$, which are left unspecified.  Results for the specific case $N_f=1$ are also presented, where the symmetries allow only five linearly independent quartic terms. For the calculation of the Green's functions, we consider both one-particle reducible and one-particle irreducible Feynman diagrams. Knowledge of these renormalization factors is necessary in order to relate numerical results, coming from nonperturbative studies, to ``physical'' observables.
\end{abstract}

\maketitle

\section{Introduction}

Supersymmetric models of strongly coupled theories hold immense promise as a frontier for exploring Beyond the Standard Model (BSM) physics. In recent years, the prospects of simulating supersymmetric extensions of Quantum Chromodynamics (QCD) on a lattice have become increasingly tangible. However, this endeavor faces several significant difficulties due to the breaking of Supersymmetry (SUSY) within the lattice framework~\cite{Dondi:1976tx, Catterall:2009it, Giedt:2009yd, Joseph:2015xwa, Bergner:2016sbv}. These challenges include the need to fine-tune the parameters of the bare Lagrangian~\cite{Elliott:2008jp, Giedt:2009yd}.

As is the case with all known regulators, lattice breaks supersymmetry explicitly but it is the only regulator which describes many aspects of strong interactions nonperturbatively. Note that the coupling constants appearing in the lattice action are not all identical. On one hand, gauge invariance, which is intact on the lattice, dictates that some of the interaction parts will have the same coupling, $g$ (gauge coupling); this is the case for the kinematic terms which contain covariant derivatives and thus gluons couple with quarks, squarks, gluinos and other gluons with the same gauge coupling constant. The Yukawa interactions between quarks, squarks and gluinos as well as the four-squark interactions contain a different coupling constant, which must be fine-tuned on the lattice. Exploiting the symmetries of the action, we reduce the number of possible interaction terms and, therefore, their tuning~\cite{Giedt:2009yd}. It is desirable to employ a lattice discretization which preserves as many as possible of the continuum symmetries, so that the relevant operators to be tuned will be fewer. The overlap formulation can be used, in order to preserve chiral symmetry, but we will first investigate these tunings using the Wilson fermion action. The use of the overlap action is beyond the scope of the present work.

Lattice studies of Supersymmetric Yang-Mills (SYM) theories have already accumulated a rich spectrum of both perturbative and non-perturbative results, significantly advancing our understanding of these theories in various dimensions, see e.g., ~\cite{Montvay:2001aj, Cabo-Bizet:2019eaf, Demmouche:2010sf, Bergner:2016sbv, Bergner:2014dua, Caetano:2016ydc, Schaich:2022xgy}. These studies have shed light on the non-perturbative dynamics of SYM theories, such as confinement, the mass spectrum of the lightest bound states, and the behavior of the theory under supersymmetry breaking. Furthermore, perturbative investigations using specific regulators, e.g.,~\cite{Bergner:2022wnb, Costa:2021pfu, Costa:2020keq, Krause:1974dj, Weinberg:1973ua, Ferrara:1975ye, Slavnov:1974dg, Taniguchi:1999fc}, have contributed to precise determinations of counterterms, fine-tuning of parameters, and the renormalization of SUSY operators. 

${\cal{N}} = 1$ Supersymmetric Quantum Chromodynamics (SQCD) serves as a prototype theory for supersymmetric extensions of the Standard Model. The study of its bound states, phase structure, and the mechanism of spontaneous Supersymmetry breaking is expected to provide insights into analogous phenomena that could occur beyond the Standard Model. This makes SQCD an important framework for exploring how Supersymmetry might manifest itself in nature and its potential impact on particle physics at higher energy scales. 

In this work, we focus on the renormalization of the quartic coupling in SQCD, which completes the fine-tuning of all parameters and fields of the SQCD Lagrangian ~\cite{Costa:2017rht,Costa:2018mvb,Costa:2023cqv}.
Our methodology involves calculations of Green's functions with four external squarks, extending up to one loop and to the lowest order in the lattice spacing. This approach enables us to renormalize the quartic couplings that appear in the SQCD action. At the same time, these Green's functions are instrumental in studying phenomena such as the supersymmetric phase transition through the analysis of the four-squark effective potential.

Our results for the renormalized quartic couplings are obtained using the Wilson gauge action for the gluon fields. For fermions (quark, gluino fields) we employ the Wilson fermion action, and for the squark fields we use na\"ive discretization. After presenting the basics of the computation setup (Section \ref{comsetUP}), we address the renormalization of the quartic couplings (Section \ref{couplingQ}) both in dimensional and lattice regularizations. We utilize the $\MSbar$ renormalization scheme and we calculate the fine-tunings for all quartic couplings to one-loop order. Finally, we end with a short outlook (Section~\ref{summary}).

\section{Formulation and Computational Setup}
\label{comsetUP} 

In the Wess-Zumino (WZ) gauge, the SQCD action contains the following fields: the gluon together with the gluino; and for each quark flavor, a Dirac fermion (quark) and two squarks. In the following we briefly define our notation. Although the action of SQCD used in this calculation can be found in literature, e.g. in Refs.~\cite{Wess:1992cp, Martin:1997ns, Costa:2017rht}, we present it here for completeness' sake; in the continuum and in Minkowski space, the action of SQCD is:
\bea
{\cal S}_{\rm SQCD} & = & \int d^4x \Big[ -\frac{1}{4}u_{\mu \nu}^{\alpha} {u^{\mu \nu}}^{\alpha} + \frac{i}{2} \bar \lambda^{\alpha}  \gamma^\mu {\cal{D}}_\mu\lambda^{\alpha}  \nonumber \\
&-& {\cal{D}}_\mu A_+^{\dagger}{\cal{D}}^\mu A_+ - {\cal{D}}_\mu A_- {\cal{D}}^\mu A_-^{\dagger}+ i \bar \psi  \gamma^\mu {\cal{D}}_\mu \psi  \nonumber \\
&-&i \sqrt2 g \big( A^{\dagger}_+ \bar{\lambda}^{\alpha}  T^{\alpha} P_+ \psi   -  \bar{\psi}  P_- \lambda^{\alpha}   T^{\alpha} A_+ +  A_- \bar{\lambda}^{\alpha}  T^{\alpha} P_- \psi   -  \bar{\psi}  P_+ \lambda^{\alpha}   T^{\alpha} A_-^{\dagger}\big)\nonumber\\  
&-& \frac{1}{2} g^2 (A^{\dagger}_+ T^{\alpha} A_+ -  A_- T^{\alpha} A^{\dagger}_-)^2 + m ( \bar \psi  \psi  - m A^{\dagger}_+ A_+  - m A_- A^{\dagger}_-)\Big] \,,
\label{susylagr}
\eea
where $\psi$ ($u_\mu$) is the quark (gluon) field, $\lambda$ is the gluino field and $A_\pm$ are the squark field components; $T^{\alpha}$ are the generators of the $SU(N_c)$ gauge group and $P_\pm$ are projectors: $P_\pm= (1 \pm \,\gamma_5)/2$. Besides an implicit color index, quark and squark fields (as well as their masses, $m$) carry also an implicit flavor index; a summation over repeated indices is intended\footnote{Note that the first parenthesis in the last line of Eq.~(\ref{susylagr}) has an implicit double summation over independent flavor indices.}. The definitions of the covariant derivatives and of the gluon field tensor are:
\bea
{\cal{D}}_\mu A_+ &=&  \pa_{\mu} A_+ + i g\,u_{\mu}^{\alpha}\,T^{\alpha}\,A_+ \nonumber \\
{\cal{D}}_\mu A_-^{\dagger} &=&  \pa_{\mu} A_-^{\dagger} + i g\,u_{\mu}^{\alpha}\,T^{\alpha}\,A_-^{\dagger} \nonumber \\
{\cal{D}}_\mu A_- &=&  \pa_{\mu} A_- - i g\,A_-\,T^{\alpha}\,u_{\mu}^{\alpha} \nonumber \\
{\cal{D}}_\mu A_+^{\dagger} &=&  \pa_{\mu} A_+^{\dagger} - i g\,A_+^{\dagger}T^{\alpha}\,u_{\mu}^{\alpha} \nonumber \\
{\cal{D}}_\mu \psi &=&  \pa_{\mu} \psi+ i g\,u_{\mu}^{\alpha} \,T^{\alpha}\,\psi \nonumber\\
{\cal{D}}_\mu \lambda &=&  \pa_{\mu} \lambda + i g \,[u_{\mu},\lambda] \nonumber\\
 u_{\mu \nu} &=& \pa_{\mu}u_{\nu} - \pa_{\nu}u_{\mu} + i g\, [u_{\mu},u_{\nu}]. 
\eea
The above action is invariant upon these supersymmetric transformations:
\bea
\delta_\xi A_+ & = & - \sqrt2  \bar\xi  P_+\psi  \, , \nonumber \\
\delta_\xi A_- & = & - \sqrt2  \bar\psi  P_+ \xi   \, , \nonumber \\
\delta_\xi (P_+ \psi) & = & i \sqrt2 ({\cal{D}}_\mu A_+) P_+ \gamma^\mu \xi   - \sqrt2 m P_+ \xi  A_-^{\dagger}\, , \nonumber \\
\delta_\xi (P_- \psi ) & = &  i \sqrt2 ({\cal{D}}_\mu A_-)^{\dagger} P_- \gamma^\mu \xi   - \sqrt2 m  A_+ P_- \xi \, ,\nonumber \\
\delta_\xi u_\mu^{\alpha} & = & -i \bar \xi  \gamma^\mu \lambda^{\alpha} , \nonumber \\
\delta_\xi \lambda^{\alpha}  & = & \frac{1}{4} u_{\mu \nu}^{\alpha} [\gamma^{\mu},\gamma^{\nu}] \xi  - i g \gamma^5 \xi  (A^{\dagger}_+ T^{\alpha} A_+ -  A_- T^{\alpha} A^{\dagger}_-)\,.
\label{susytransfDirac}
\eea

As in the case with the quantization of ordinary gauge theories, additional infinities will appear upon functionally integrating over gauge orbits. The standard remedy is to introduce a gauge-fixing term in the Lagrangian, along with a compensating Faddeev-Popov ghost term. The resulting Lagrangian, though no longer manifestly gauge invariant, is still invariant under Becchi-Rouet-Stora-Tyutin (BRST) transformations. This procedure of gauge fixing guarantees that Green's functions of gauge invariant objects will be gauge independent to all orders in perturbation theory. We use the ordinary gauge fixing term and ghost contribution arising from the Faddeev-Popov gauge fixing procedure:
\begin{equation}
{\cal S}^E_{GF}= \frac{1}{\alpha}\int d^4x {\rm{Tr}} \left( \partial_\mu u_\mu\right)^2,
\label{sgf}
\end{equation}
where $\alpha$ is the gauge parameter [$\alpha=1(0)$ corresponds to Feynman (Landau) gauge], and 
\begin{equation}
{\cal S}^E_{Ghost}= - 2 \int d^4x {\rm{Tr}} \left( \bar{c}\, \partial_{\mu}D_\mu  c\right), 
\label{sghost}
\end{equation}
where the ghost field $c$ is a Grassmann scalar which transforms in the adjoint representation of the gauge group, and: ${\cal{D}}_\mu c =  \pa_{\mu} c + i g \,[u_\mu,c]$. This gauge fixing term breaks supersymmetry. However, given that the renormalized theory does not depend on the choice of a gauge fixing term, and given that all known regularizations, in particular the lattice regularization, violate supersymmetry at intermediate steps, one may choose this standard covariant gauge fixing term.

In Refs.~\cite{Costa:2017rht} and~\cite{Costa:2018mvb}, first lattice perturbative computations in the context of SQCD were presented; there, apart from the Yukawa and quartic couplings, we extracted the renormalization of all parameters and fields appearing in Eq.~(\ref{susylagr}) using Wilson gluons and fermions. In addition, we explored the mixing of some composite operators under renormalization. The results in these references~\cite{Costa:2017rht,Costa:2018mvb} will find further use in the present work. In this article we calculate the fine-tunings of the 4-squark quartic couplings, which are an essential prerequisite towards non-perturbative investigations. Note that we have also calculated the fine-tunings of the Yukawa couplings in Ref.~\cite{Costa:2023cqv}.

From this point on, we switch to Euclidean space. In our lattice calculation, we extend Wilson's formulation of the QCD action, to encompass SUSY partner fields as well. In this standard discretization quarks, squarks and gluinos are defined on the lattice sites, while gluons are defined on the links of the lattice: $U_\mu (x) = e^{i g a T^{\alpha} u_\mu^\alpha (x+a\hat{\mu}/2)}$; $\alpha$ is a color index in the adjoint representation of the gauge group. This formulation leaves no SUSY generators intact, and it also breaks chiral symmetry; thus, the need for fine-tuning will arise in numerical simulations of SQCD. For Wilson-type quarks and gluinos, the Euclidean action ${\cal S}^{L}_{\rm SQCD}$ on the lattice becomes:       
\bea
{\cal S}^{L}_{\rm SQCD} & = & a^4 \sum_x \Big[ \frac{N_c}{g^2} \sum_{\mu,\,\nu}\left(1-\frac{1}{N_c}\, {\rm Tr} U_{\mu\,\nu} \right ) + \sum_{\mu} {\rm Tr} \left(\bar \lambda  \gamma_\mu {\cal{D}}_\mu\lambda  \right ) - a \frac{r}{2} {\rm Tr}\left(\bar \lambda   {\cal{D}}^2 \lambda  \right) \nonumber \\ 
&+&\sum_{\mu}\left( {\cal{D}}_\mu A_+^{\dagger}{\cal{D}}_\mu A_+ + {\cal{D}}_\mu A_- {\cal{D}}_\mu A_-^{\dagger}+ \bar \psi  \gamma_\mu {\cal{D}}_\mu \psi  \right) - a \frac{r}{2} \bar \psi   {\cal{D}}^2 \psi  \nonumber \\
&+&i \sqrt2 g \big( A^{\dagger}_+ \bar{\lambda}^{\alpha}  T^{\alpha} P_+ \psi   -  \bar{\psi}  P_- \lambda^{\alpha}   T^{\alpha} A_+ +  A_- \bar{\lambda}^{\alpha}  T^{\alpha} P_- \psi   -  \bar{\psi}  P_+ \lambda^{\alpha}   T^{\alpha} A_-^{\dagger}\big)\nonumber\\  
&+& \frac{1}{2} g^2 (A^{\dagger}_+ T^{\alpha} A_+ -  A_- T^{\alpha} A^{\dagger}_-)^2 - m ( \bar \psi  \psi  - m A^{\dagger}_+ A_+  - m A_- A^{\dagger}_-)
\Big] \,,
\label{susylagrLattice}
\eea
where: $a$ is the lattice spacing, $U_{\mu \nu}(x) =U_\mu(x)U_\nu(x+a\hat\mu)U^\dagger_\mu(x+a\hat\nu)U_\nu^\dagger(x)$, and a summation over flavors is understood in the last three lines of Eq.~(\ref{susylagrLattice}). The 4-vector $x$ is restricted to the values $x = na$, with $n$ being an integer 4-vector. The terms proportional to the Wilson parameter, $r$, eliminate the problem of fermion doubling, at the expense of breaking chiral invariance. In the limit $a \to 0$ the lattice action reproduces the continuum one. The bare couplings for the Yukawa and quartic terms (last two lines of Eq.\eqref{susylagrLattice}) need not coincide with the gauge coupling $g$; this requirement is imposed on the respective renormalized values.

The definitions of the covariant derivatives are as follows:
\bea
{\cal{D}}_\mu\lambda (x) &\equiv& \frac{1}{2a} \Big[ U_\mu (x) \lambda  (x + a \hat{\mu}) U_\mu^\dagger (x) - U_\mu^\dagger (x - a \hat{\mu}) \lambda  (x - a \hat{\mu}) U_\mu(x - a \hat{\mu}) \Big] \\
{\cal D}^2 \lambda (x) &\equiv& \frac{1}{a^2} \sum_\mu \Big[ U_\mu (x)  \lambda  (x + a \hat{\mu}) U_\mu^\dagger (x)  - 2 \lambda (x) +  U_\mu^\dagger (x - a \hat{\mu}) \lambda  (x - a \hat{\mu}) U_\mu(x - a \hat{\mu})\Big]\\
{\cal{D}}_\mu \psi (x) &\equiv& \frac{1}{2a}\Big[ U_\mu (x) \psi  (x + a \hat{\mu})  - U_\mu^\dagger (x - a \hat{\mu}) \psi  (x - a \hat{\mu})\Big]\\  
{\cal D}^2 \psi (x) &\equiv& \frac{1}{a^2} \sum_\mu \Big[U_\mu (x) \psi  (x + a \hat{\mu})  - 2 \psi (x) +  U_\mu^\dagger (x - a \hat{\mu}) \psi  (x - a \hat{\mu})\Big]\\
\label{DAplus}
{\cal{D}}_\mu A_+(x) &\equiv& \frac{1}{a} \Big[  U_\mu (x) A_+(x + a \hat{\mu}) - A_+(x)   \Big]\\
{\cal{D}}_\mu A_+^{\dagger}(x) &\equiv& \frac{1}{a} \Big[A_+^{\dagger}(x + a \hat{\mu}) U_\mu^{\dagger}(x)  -  A_+^\dagger(x)\Big]\\
{\cal{D}}_\mu A_-(x) &\equiv& \frac{1}{a} \Big[A_-(x + a \hat{\mu}) U_\mu^{\dagger}(x)  -  A_-(x)\Big]\\
\label{DAminusdagger}
{\cal{D}}_\mu A_-^{\dagger}(x) &\equiv& \frac{1}{a} \Big[U_\mu (x) A_-^{\dagger}(x + a \hat{\mu})   -  A_-^{\dagger}(x) \Big]
\eea
Note that in Eqs.~(\ref{DAplus})-(\ref{DAminusdagger}), in order not to involve more that two lattice points, we do not use the symmetric derivative.

By analogy to the continuum case, a discrete version of a gauge-fixing term, together with the compensating ghost field term, must be added to the action, in order to avoid divergences from the integration over gauge orbits; these terms are the same as in the non-supersymmetric case. Although these terms can be found in literature, we present them here for the sake of completeness:: 
\be
S_{GF}^{L} = \frac{1}{2\alpha} a^2 \sum_x \sum_\mu{\rm{Tr}} \left(u_\mu(x + a \hat{\mu}/2)-u_\mu(x - a \hat{\mu}/2)\right)^2.
\ee
\bea
S_{Ghost}^{L} &=& 2a^2 \sum_x\sum_\mu {\rm{Tr}} \{(\bar c(x + a \hat{\mu}) - \bar c(x))( c(x + a \hat{\mu}) - c(x) \\\nonumber
&&+ ig [u_\mu (x+ a \hat{\mu}/2),c(x)] + \frac{1}{2}ig [u_\mu (x+ a \hat{\mu}/2),c(x+a \hat{\mu}) -c(x)]\\\nonumber
&&-\frac{1}{12}g^2[u_\mu (x+ a \hat{\mu}/2),[u_\mu (x+ a \hat{\mu}/2),c(x+a \hat{\mu}) -c(x)]])\}+{\cal{O}}(g^3).
\eea
Similarly, a standard ``measure'' term must be added to the action, in order to account for the Jacobian in the change of integration variables: $U_\mu \to u_\mu$\,:
\be
S_{M}^{L} = \frac{g^2N_c}{12}a^2\sum_x \sum_\mu {\rm{Tr}}\left(u_\mu(x+ a \hat{\mu}/2)^2\right) + {\cal{O}}(g^4).
\ee

Parity $\cal{P}$ and Charge Conjugation $\mathcal {C}$ are symmetries of the lattice action and their definitions are presented below.
\be
{\cal{P}}:\left \{\begin{array}{ll}
&\hspace{-.3cm} U_0(x)\rightarrow U_0(x_{\cal{P}})\, ,\qquad U_k(x)\rightarrow U_k^{\dagger}(x_{\cal{P}}-a\hat{k})\, ,\qquad k=1,2,3\\[4pt]
&\hspace{-.3cm} \psi(x)\rightarrow \gamma_0  \psi(x_{\cal{P}})\\[4pt]
&\hspace{-.3cm}\bar{ \psi}(x) \rightarrow\bar{ \psi}(x_{\cal{P}})\gamma_0\\[4pt]
&\hspace{-.3cm} \la^{\alpha}(x) \rightarrow \gamma_0  \la^{\alpha}(x_{\cal{P}})\\[4pt]
&\hspace{-.3cm}\bar{ \la}^{\alpha}(x) \rightarrow\bar{ \la}^{\alpha}(x_{\cal{P}})\gamma_0\\[4pt]
&\hspace{-.3cm} A_\pm(x) \rightarrow A_\mp^\dagger(x_{\cal{P}})\\[4pt]
&\hspace{-.3cm} A_\pm^\dagger(x) \rightarrow A_\mp(x_{\cal{P}})
\end{array}\right .
\label{Parity}
\ee
where $x_{\cal{P}}=(-{\bf{x}},x_0)$.

\be
{\mathcal {C}}:\left \{\begin{array}{ll}
&\hspace{-.3cm}U_\mu(x)\rightarrow U_\mu^{\star}(x)\, ,\quad \mu=0,1,2,3\\[4pt]
&\hspace{-.3cm}\psi(x)\rightarrow -C \bar{ \psi}(x)^{T}\\[4pt]
&\hspace{-.3cm}\bar{\psi}(x)\rightarrow{\psi}(x)^{T}C^{\dagger}\\[4pt]
&\hspace{-.3cm} \la(x) \rightarrow C \bar{\la}(x)^{T}\\[4pt]
&\hspace{-.3cm}\bar{\la}(x) \rightarrow -{\la}(x)^{T}C^{\dagger}\\[4pt]
&\hspace{-.3cm}A_\pm(x) \rightarrow A_\mp(x) \\[4pt]
&\hspace{-.3cm}A_\pm^\dagger(x) \rightarrow A_\mp^\dagger(x)
\end{array}\right . 
\label{Chargeconjugation}
\ee
where $^{\,T}$ means transpose (also in the $SU(N_c)$ generators implicit in the gluino fields). The matrix $C$ satisfies: $(C \gamma_{\mu})^{T}= C \gamma_{\mu}$, $C^T=-C$ and $C^{\dagger} C=1$. In 4 dimensions, in a standard basis for $\gamma$ matrices, in which $\gamma_0,\ \gamma_2$ ($\gamma_1,\ \gamma_3$) are (anti-)symmetric, $C = - {\rm i} \gamma_0 \gamma_2$. Note that all operators that we consider here are flavor singlets. 

Further symmetries in the massless case are: 

$U(1)_R$ rotates the quark and gluino fields in opposite direction:
\be
{\cal{R}}:\left \{\begin{array}{ll}
&\hspace{-.3cm} \psi(x)\rightarrow e^{i \theta \gamma_5}  \psi(x)\\[4pt]
&\hspace{-.3cm}\bar{ \psi}(x) \rightarrow\bar{ \psi}(x)e^{i \theta \gamma_5}\\[4pt]
&\hspace{-.3cm} \la(x)\rightarrow e^{-i \theta \gamma_5}  \la(x)\\[4pt]
&\hspace{-.3cm}\bar{ \la}(x) \rightarrow\bar{ \la}(x)e^{-i \theta \gamma_5}
\end{array}\right.
\label{Rsym}
\ee
The ${\cal{R}}$-symmetry does not commute with the SUSY generators.  

$U(1)_A$ rotates the squark and the quark fields in the same direction as follows: 
\be
{\cal{\chi}}:\left \{\begin{array}{ll}
&\hspace{-.3cm} \psi(x)\rightarrow e^{i \theta' \gamma_5}  \psi(x)\\[4pt]
&\hspace{-.3cm}\bar{ \psi}(x) \rightarrow\bar{ \psi}(x)e^{i \theta' \gamma_5}\\[4pt]
&\hspace{-.3cm} A_\pm(x) \rightarrow e^{ i \theta'} A_\pm(x)\\[4pt]
&\hspace{-.3cm} A_\pm^\dagger(x) \rightarrow e^{- i \theta'} A_\pm^\dagger(x)
\end{array}\right .
\label{chiral}
\ee

We compute, perturbatively, the relevant four-point (4-pt) Green's functions using both dimensional regularization ($DR$) in $D = 4 - 2\epsilon$ dimensions and lattice regularization ($LR$). These one-loop computations are the crux of this paper and sheer difficulties involving SQCD action and its discretization as well as four-point Green's functions are quite complicated. Combining the results for the bare Green's functions on the lattice and the renormalized Green's functions found in the continuum, we are able to extract the renormalization factors appropriate to lattice regularization and $\MSbar$ renormalization scheme. The renormalizations of the gluon and squark fields and the gauge coupling are a prerequisite for the renormalization of the quartic couplings, since renormalization conditions in vertex corrections involve these quantities.

The 4-point Green's functions involve four external squark fields in momentum space. For completeness, we also present our conventions for Fourier transformations: 
\begin{align}
\tilde \psi(q) &= \int d^4x\, e^{-i q \cdot x}\,\psi(x) \\
\tilde A_\pm(q) &= \int d^4x\, e^{\mp i q \cdot x}\,A_\pm(x)\\
\tilde{u}_\mu(q) &= \int d^4x\, e^{-i q \cdot x}\,u_\mu(x)\\%
\tilde \la(q) &= \int d^4x\, e^{-i q \cdot x}\,\la(x)
\label{Fourier}
\end{align}  
To avoid heavy notation, we will omit the tilde from the Fourier-transformed fields. These fields will be understood from their arguments. 

There exist several prescriptions~\cite{Larin:1993tq} for defining $\gamma_5$ in $D$ dimensions, such as the na\"ive dimensional regularization (NDR)~\cite{Chanowitz:1979zu}, the t'Hooft-Veltman (HV)~\cite{tHooft:1972}, the $DRED$~\cite{Siegel:1979} and the $DR{\overline{EZ}}$ prescriptions (see, e.g., Ref.~\cite{Patel}). They are related among themselves via finite conversion factors~\cite{Buras:1989xd}. In our calculation, we apply the NDR and HV prescriptions. The latter does not violate Ward identities involving pseudoscalar and axial-vector operators in $D$ dimensions. The metric tensor, $\eta_{\mu\nu}$, and the Dirac matrices, $\gamma_\mu$, satisfy the following relations in $D$ dimensions:
\be
\eta^{\mu\nu}\eta_{\mu\nu}=D,\qquad \{\gamma_\mu,\gamma_\nu\} = 2 \delta_{\mu\nu} \openone.
\ee
In NDR, the definition of $\gamma_5$ satisfies:
\be
\{\gamma_5,\gamma_{\mu}\} = 0, \, \,\forall \mu,
\ee
whereas in HV it satisfies:
\be
\{\gamma_5,\gamma_{\mu}\} = 0, \, \,\mu = 1,2,3,4, \qquad [\gamma_5,\gamma_{\mu}]=0, \,\, \mu>4.
\ee

\section{Renormalization of the quartic coupling}
\label{couplingQ}
In this section, we present our one-loop calculations for the bare 4-point Green's functions and the renormalization factors of the quartic couplings in the $\MSbar$ scheme, employing both dimensional regularization ($DR$) and lattice regularization ($LR$). To renormalize the quartic couplings, we impose specific renormalization conditions, ensuring the cancellation of divergences in the corresponding bare 4-point amputated Green's functions with four external squark fields. Applying these renormalization factors to the bare Green's functions yields the renormalized Green's functions, which are independent of the regulator ($\epsilon$ in DR, $a$ in LR).

Quartic couplings (four-squark interactions) must be appropriately fine-tuned on the lattice. The gauge symmetry allows two squarks to lie in the fundamental representation and the other two in the antifundamental; ignoring flavor indices for the moment, there are ten possible types of counterterms related to quartic couplings:
\bea
\label{externalSs}
&& (A_+^{\dagger}  A_+)( A_+^{\dagger}  A_+ ), 
\quad  (A_- A_-^{\dagger})( A_- A_-^{\dagger}), \\\nonumber &&
(A_+^{\dagger}  A_+) (A_- A_-^{\dagger}),
\quad (A_+^{\dagger}  A_-^{\dagger} )(A_+^{\dagger}  A_-^{\dagger}), 
\quad  (A_- A_+ )(A_- A_+), \quad (A_- A_+)(A_+^{\dagger}  A_-^{\dagger}),  \\\nonumber &&
(A_+^{\dagger}  A_+)( A_+^{\dagger}  A_-^{\dagger}),
\quad  (A_+^{\dagger}  A_+)( A_- A_+),  
\quad  (A_- A_-^{\dagger})( A_+^{\dagger}  A_-^{\dagger}), 
\quad  (A_- A_-^{\dagger})( A_- A_+ )
\eea
Pairs of squark fields in parentheses denote color-singlet combinations. We will be calculating a total of 9 Green's functions containing four squarks, namely those for which each of the 10 potential counterterms of Eq.~(\ref{externalSs}) has a nonzero lowest order contribution. [Note that $(A_+^{\dagger}  A_+) (A_- A_-^{\dagger})$ and $(A_- A_+)(A_+^{\dagger}  A_-^{\dagger})$ contribute to the same Green's function.] Note that the first three terms from Eq.~(\ref{externalSs}) appear in the SQCD action and they will be used to obtain the renormalization factor of the quartic coupling. The other terms can (and will) appear as counterterms, in combinations which are invariant under all symmetries of the lattice SQCD action, including $\cal{C}$ and $\cal{P}$ symmetries.

For the case where the flavor number is $N_f=1$, there are only five possible combinations \cite{Wellegehausen:2018opt}, which are the first five terms in Table~\ref{tb:nonsinglet2}; compared to the terms present in the SQCD action, there are four more terms ($\la_2, \la_3, \la_4, \la_5 $). However, in this study, we consider $N_f$ fundamental multiplets, and the flavor symmetries allow for an additional five ``Fierz-transform'' operators to be fine-tuned. To clarify, we provide an example of an operator with $N_f>1$   that can and will contribute to the fine-tuning on the lattice:
\begin{align*}
\mathcal{O} &= \sum_{c,c',f,f'} \left(A^c_{-f} A^{c\dagger}_{-f}\right)\left( A^{c'}_{-f'} A^{c'}_{+f'} \right) \\
\mathcal{O}^F &= \sum_{c,c',f,f'} \left(A^c_{-f} A^{c\dagger}_{-f'}\right)\left( A^{c'}_{-f'} A^{c'}_{+f} \right),
\end{align*}
where the superscript letter $F$ stands for Fierz, the flavor and color indices are denoted by Latin letters $f,f'$ and $c,c'$, respectively. Since $N_f$ remains unspecified throughout our calculations, there are ten possible combinations of quartic squark terms, as shown in Table~\ref{tb:nonsinglet2}. 

\begin{table}[ht!]
\begin{center}
\scalebox{1}{
  \begin{tabular}{c|c|c}
\hline \hline
Operators & $\cal{C}$&$\cal{P}$ \\ [0.5ex] \hline \hline
    $\la_1 \left[(A_{+ \, f}^{\dagger} \, T^{\alpha} \, A_{+ \, f})(A_{+ \, f'}^{\dagger} \, T^{\alpha} \, A_{+ \, f'}) - 2 (A_{+ \, f}^{\dagger} \, T^{\alpha} \, A_{+ \, f}) (A_{-\,f'} \, T^{\alpha} \, A_{-\,f'}^{\dagger})  + (A_{-\,f} \, T^{\alpha} \, A_{-\,f}^{\dagger})(A_{-\,f'} \, T^{\alpha} \, A_{-\,f'}^{\dagger}) \right]/2 $&$+$& $+$   \\[0.75ex]\hline
    $\la_2 \left[(A_{+ \, f}^{\dagger}  A_{-\,f}^{\dagger})(A_{+\,f'}^{\dagger}  A_{-\,f'}^{\dagger})+ (A_{-\,f} A_{+ \, f})(A_{-\,f'} A_{+ \, f'})\right] $&$+$& $+$   \\[0.75ex]\hline
    $\la_3  (A_{+ \, f}^{\dagger}  A_{+ \, f})  (A_{-\,f'} A_{-\,f'}^{\dagger}) $&$+$ & $+$  \\[0.75ex]\hline
    $\la_4 (A_{+ \, f}^{\dagger}  A_{-\,f}^{\dagger}) (A_{-\,f'} A_{+ \, f'}) $&$+$& $+$   \\[0.75ex]\hline
    $\la_5 (A_{+ \, f}^{\dagger}  A_{-\,f}^{\dagger} + A_{-\,f} A_{+ \, f})(A_{+ \, f'}^{\dagger}  A_{+ \, f'}  + A_{-\,f'} A_{-\,f'}^{\dagger}) $&$+$ & $+$  \\[0.75ex]\hline
    $\la_1^F \left[(A_{+ \, f}^{\dagger} \, T^{\alpha} \, A_{+ \, f'})(A_{+ \, f'}^{\dagger} \, T^{\alpha} \, A_{+ \, f}) - 2 (A_{+ \, f}^{\dagger} \, T^{\alpha} \, A_{+ \, f'}) (A_{-\,f'} \, T^{\alpha} \, A_{-\,f}^{\dagger})  + (A_{-\,f} \, T^{\alpha} \, A_{-\,f'}^{\dagger})(A_{-\,f'} \, T^{\alpha} \, A_{-\,f}^{\dagger}) \right]/2 $&$+$& $+$   \\[0.75ex]\hline
    $\la_2^F \left[(A_{+ \, f}^{\dagger}  A_{-\,f'}^{\dagger})(A_{+\,f'}^{\dagger}  A_{-\,f}^{\dagger})+ (A_{-\,f} A_{+ \, f'})(A_{-\,f'} A_{+ \, f})\right] $&$+$& $+$   \\[0.75ex]\hline
    $\la_3^F  (A_{+ \, f}^{\dagger}  A_{+ \, f'})  (A_{-\,f'} A_{-\,f}^{\dagger}) $&$+$ & $+$  \\[0.75ex]\hline
    $\la_4^F (A_{+ \, f}^{\dagger}  A_{-\,f'}^{\dagger}) (A_{-\,f'} A_{+ \, f}) $&$+$& $+$   \\[0.75ex]\hline
    $\la_5^F (A_{+ \, f}^{\dagger}  A_{-\,f'}^{\dagger} + A_{-\,f} A_{+ \, f'})(A_{+ \, f'}^{\dagger}  A_{+ \, f}  + A_{-\,f'} A_{-\,f}^{\dagger}) $&$+$ & $+$  \\[0.75ex]\hline
\hline
\end{tabular}}
\caption{Dimension-4 operators which are gauge invariant and flavor singlets. All operators appearing in this table are eigenstates of charge conjugation, $\cal{C}$, and parity, $\cal{P}$, with eigenvalue 1. The flavor indices ($f$, $f'$) on the squark fields are explicitly displayed, while the color indices, which are the same within each parenthesis, are implicit; a summation is implied over all flavor and color indices. The parameters $\la_i$ and $\la_i^F$ denote the ten quartic couplings.}
\label{tb:nonsinglet2}
\end{center}
\end{table}

Making use of the transformation ${\cal \chi} \times {\cal R}$ for the combinations shown in Table~\ref{tb:nonsinglet2}, $(A_+^{\dagger}  A_-^{\dagger})^2 + (A_- A_+)^2$ and $(A_+^{\dagger}  A_-^{\dagger} + A_- A_+)(A_+^{\dagger}  A_+  + A_- A_-^{\dagger})$ are not invariant; however, they may appear in our one-loop computations, exhibiting an anomalous behavior of this transformation. 

The first combination in Table~\ref{tb:nonsinglet2} aligns with the first term of the fourth line of Eq.~(\ref{susylagrLattice}). However, at the quantum level, the other combinations may emerge, having a potentially different quartic coupling, denoted as $\la_i$. In the classical continuum limit, $\la_1$ corresponds to $g^2$, while $\la_{2-5}$ and $\la^F_{1-5}$ vanish; thus, the tree-level values of $\la_i$ (quartic couplings) which satify SUSY are:
\be
\la_1 =  g^2,  \, \, \, \la_2 = \la_3 = \la_4= \la_5 = \la^F_1 = \la^F_2 = \la^F_3 = \la^F_4= \la^F_5 = 0 \, .
\ee

\begin{figure}[ht!]
\centering
\includegraphics[scale=0.32]{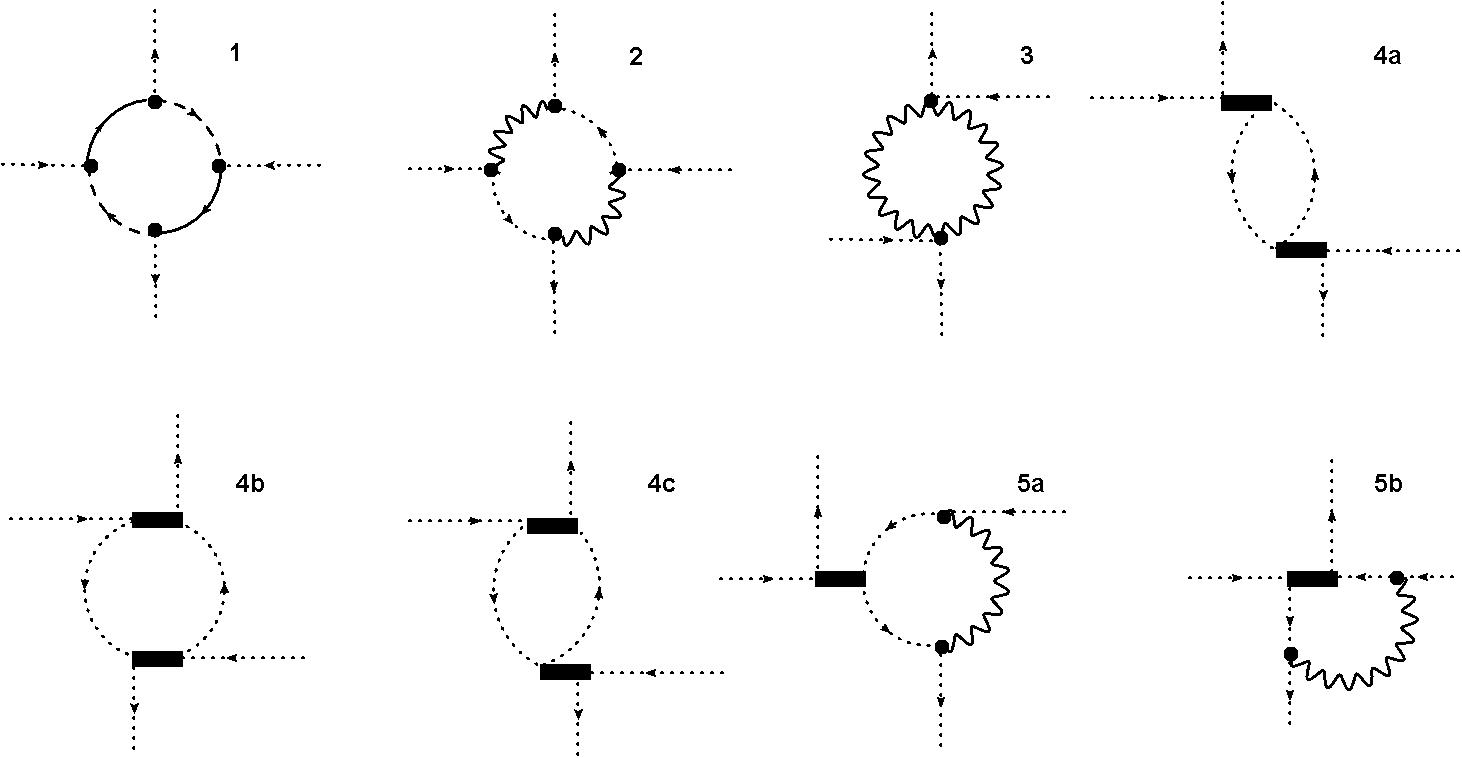} \\
\vspace{1cm}
\hspace{1cm} \includegraphics[scale=0.32]{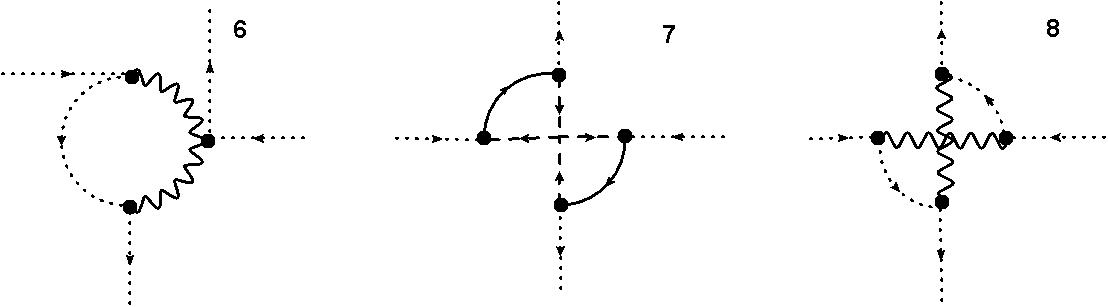}\\
\vspace{1cm}
\hspace{2cm}
\caption{One-loop 1PI Feynman diagrams leading to the fine-tuning of the quartic couplings.  A wavy (solid) line represents gluons (quarks). A dotted (dashed) line corresponds to squarks (gluinos). 
In the above diagrams the directions of the external line depend on the particular Green's function under study. An arrow entering (exiting) a vertex denotes a $\la, \psi, A_+, A_-^{\dagger}$ ($\bar \la, \bar \psi, A_+^{\dagger}, A_-$) field. The 4-squark vertex of the action has been denoted by a solid rectangle, in order to indicate the squark-antisquark color pairing; all remaining vertices are denoted by a solid circle. Squark lines could be further marked with a $+$($-$) sign, to denote an $A_+ \, (A_-)$ field.  All diagrams can have mirror variants. In diagrams 4 and 5, there are additional variants in which two external outgoing (or incoming) lines stem from a 4-squark vertex. }
\label{quarticFD1PI}
\end{figure}

\begin{figure}[ht!]
\centering
\includegraphics[scale=0.32]{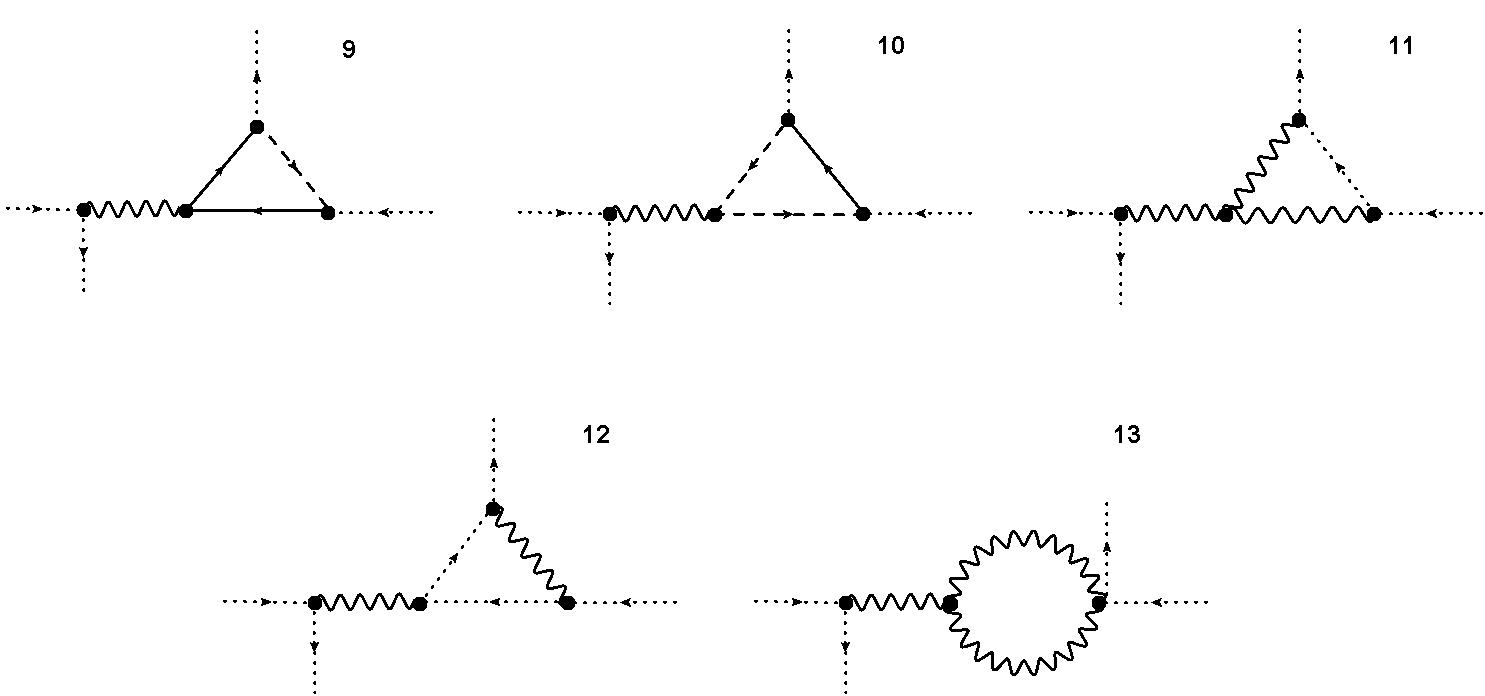}
\newline
\newline
\newline
\includegraphics[scale=0.32]{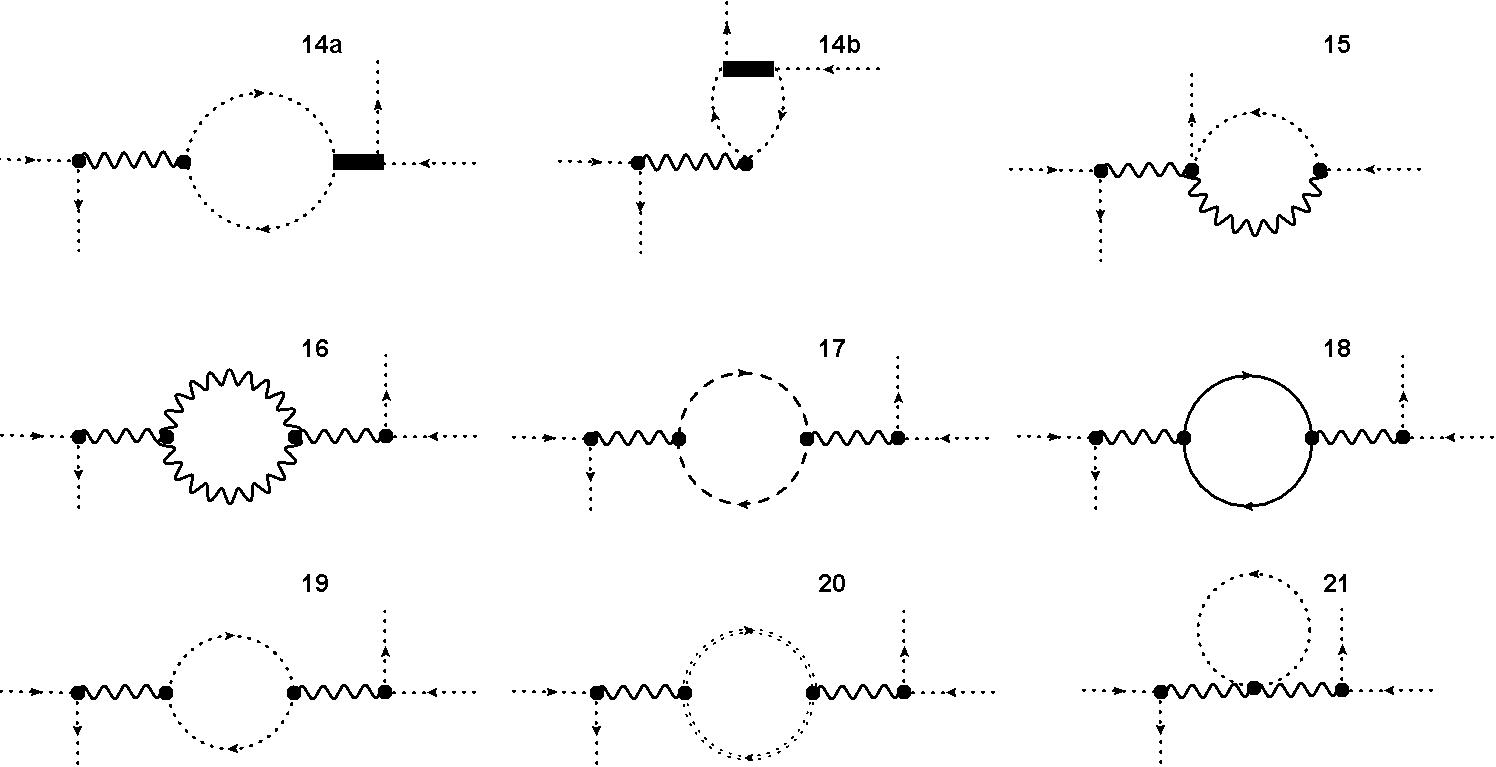}
\caption{One-loop 1PR Feynman diagrams leading to the fine-tuning of the quartic couplings.  Notation is identical to that of Figure \ref{quarticFD1PI}.  Note that the “double dashed” line is the ghost field. All diagrams can have mirror variants. Unlike gluon tadpoles which vanish in dimensional regularization, the massive squark tadpole gives a nonzero contribution (diagram 21). 
}
\label{quarticFD1PR}
\end{figure}

These couplings receive quantum corrections, coming from the Feynman diagrams of Figs.~\ref{quarticFD1PI} and \ref{quarticFD1PR}. Note that here we introduce a non-zero mass for quarks and squarks in order to avoid IR divergences. It is worth mentioning that the Majorana nature of gluinos manifests itself in some diagrams, in which $\lambda-\lambda$ as well as $\bar{\lambda}-\bar{\lambda}$ propagators appear. The Majorana condition is the following:
\begin{equation}
    (\bar \la^{\alpha})^T= C\la^{\alpha} \, ,
\end{equation}
and the tree-level propagators that relate $\lambda-\lambda$ and $\bar{\lambda}-\bar{\lambda}$ are:
\begin{align}
    \langle \lambda^{\alpha_1}(q_1) \lambda^{\alpha_2}(q_2) \rangle^{\rm{tree}} &= 2 i \, \delta^{\alpha_1 \alpha_2} \delta(q_1+q_2)  \frac{1}{\qslash_1} C^{\dagger} \\
    \langle \bar{\lambda}^{\alpha_1}(q_1) \bar{\lambda}^{\alpha_2}(q_2) \rangle^{\rm{tree}} &= - 2 i \, C \, \delta^{\alpha_1 \alpha_2} \delta(q_1+q_2)  \frac{1}{\qslash_2} \, .
\end{align}

As in the case of the Yukawa coupling~\cite{Costa:2023cqv}, for convenience of computation, we are free to make appropriate choices of the external momenta in 4-point Green's functions. Having checked that no superficial IR divergences will be generated, we will compute the diagrams by setting to zero the momenta of the two external squark fields in the fundamental representation. This choice of exceptional external momenta is allowed provided that the matter fields are massive; such a requirement would not be necessary for four-point Green's functions with generic external momenta, to one-loop order. Furthermore, these four-point Green's functions will be symmetric with respect to the exchange of identical external fields.

\begin{figure}[ht!]
\centering
\includegraphics[scale=0.32]{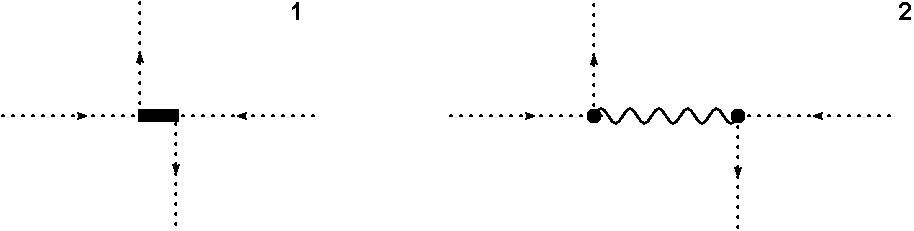}
\caption{Tree-level Feynman diagrams with four external squark fields. The second diagram can have mirror variants. A wavy line represents gluons and a dotted line corresponds to squarks. 
Squark lines could be further marked with a $+$($-$) sign, to denote an $A_+ \, (A_-)$ field. The 4-squark vertex of the action has been denoted by a solid rectangle, in order to indicate the squark-antisquark color pairing; all remaining vertices are denoted by a solid circle. An arrow entering (exiting) a vertex denotes a $A_+, A_-^{\dagger}$ ($A_+^{\dagger}, A_-$) field.
}
\label{quarticTL}
\end{figure}

\begin{figure}[ht!]
\centering
\includegraphics[scale=0.32]{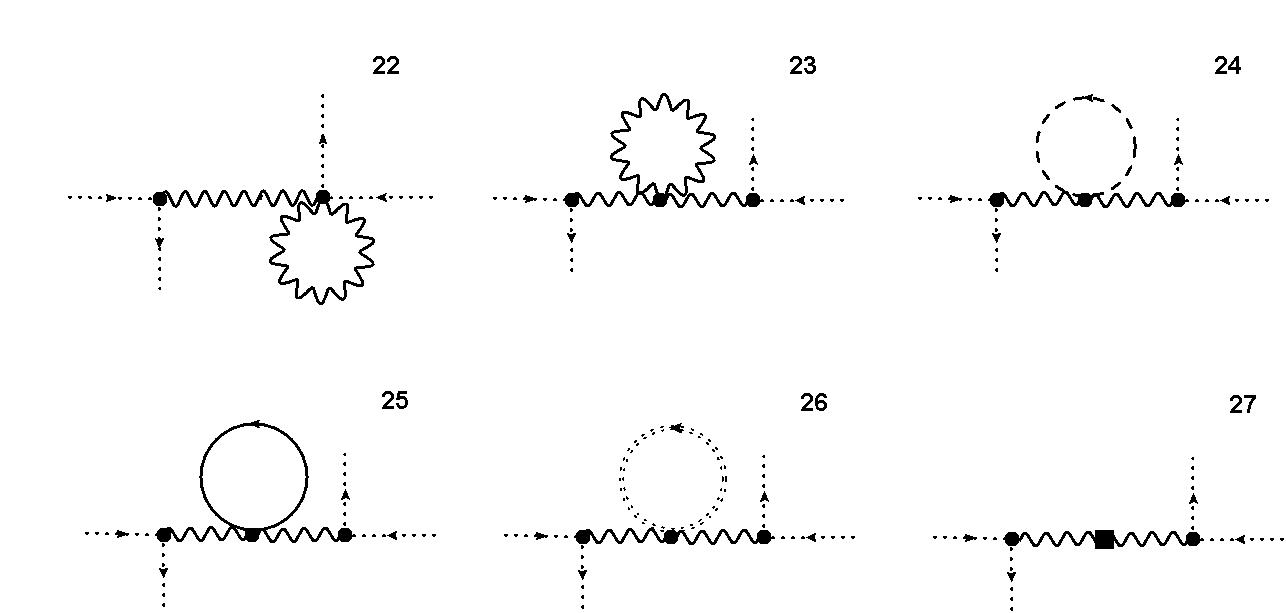}
\caption{Additional one-loop Feynman diagrams leading to the fine-tuning of the quartic couplings on the lattice. Notation is identical to that of Fig. \ref{quarticFD1PI}. Note that the “double dashed” line is the ghost field and the solid box in diagram 27 comes from the measure part of the lattice action.
}
\label{quarticFD}
\end{figure}

Let us first present the tree-level Green's functions, whose Feynman diagrams are shown in Fig.~\ref{quarticTL}, with four external squarks\footnote{In all 4-point Green's functions presented below an overall factor of $(2\pi)^4 \delta(\pm q_1 \pm q_2 \pm q_3 \pm q_4)$ is understood; the sign in front of each momentum is $+$ if it corresponds to an $A_+^\dagger$ or $A_-$ field, and it is $-$ otherwise.}:
\bea
\langle  A_{+ \, f_1}^{\dagger \alpha_1}(q_1)  A_{+ \, f_2}^{\dagger \alpha_2}(q_2) \, A_{+ \, f_3}^{\alpha_3}(q_3) A_{+ \, f_4}^{\alpha_4}(q_4) \rangle^{\rm{tree}}  &=& \langle  A_{- \, f_1}^{\dagger \alpha_1}(q_1)  A_{- \, f_2}^{\dagger \alpha_2}(q_2) \, A_{- \, f_3}^{\alpha_3}(q_3) A_{- \, f_4}^{\alpha_4}(q_4) \rangle^{\rm{tree}} \nonumber \\
&=& \frac{1}{2 N_c} \,  g^2 \, (-1 - \alpha) \times \nonumber \\
& \bigg [ &\delta_{f_1 f_3} \delta_{f_2 f_4}(-\delta^{\alpha_1 \alpha_3} \delta^{\alpha_2 \alpha_4}+N_c \, \delta^{\alpha_1 \alpha_4} \delta^{\alpha_2 \alpha_3}) \nonumber \\
&+& \delta_{f_1 f_4} \delta_{f_2 f_3}(-\delta^{\alpha_1 \alpha_4} \delta^{\alpha_2 \alpha_3}+N_c \, \delta^{\alpha_1 \alpha_3} \delta^{\alpha_2 \alpha_4}) \bigg ]\\
\langle  A_{+ \, f_1}^{\dagger \alpha_1}(q_1)  A_{+ \, f_2}^{\alpha_2}(q_2) \, A_{- \, f_3}^{\dagger \alpha_3}(q_3) A_{- \, f_4}^{\alpha_4}(q_4) \rangle^{\rm{tree}}  &=& \frac{1}{2 N_c} \, g^2 \, (1 -\alpha) \, \delta_{f_1 f_2} \delta_{f_4 f_3} \,(N_c \delta^{\alpha_1 \alpha_3} \delta^{\alpha_4 \alpha_2} - \delta^{\alpha_1 \alpha_2} \delta^{\alpha_4 \alpha_3}) 
\eea
where $f_i$ are the flavour indices of the external squark fields. The rest of the tree-level Green's functions with four external squarks are zero. In the calculation of Green's functions, both one-particle irreducible (1PI) and one-particle reducible (1PR) Feynman diagrams are considered. The one-loop diagrams in the continuum are shown in Figs.~\ref{quarticFD1PI} and~\ref{quarticFD1PR} respectively; further diagrams contributing on the lattice are shown in Fig.~\ref{quarticFD}.

In order to obtain the renormalized quartic couplings, we impose renormalization conditions which require the cancellation of divergences present in the corresponding bare 4-pt Green's functions with external squark fields and thus, the renormalization factors are defined in such a way as to remove all divergences. The application of the renormalization factors on the bare Green's functions leads to the renormalized Green's functions, which are independent of the regulator ($\epsilon$ in $DR$, $a$ in $LR$). 

The choice of the external momenta for Green's functions will not affect their pole parts in $DR$ or their logarithmic dependence on the lattice spacing in $LR$.
Since the difference between the $\MSbar$-renormalized and the corresponding Green's function enters in the extraction of the one-loop renormalization of the quartic couplings, we present below this difference. Since this difference must be polynomial in the external fields (of 0th order for the Green's function of interest), it becomes apparent that our simplifying choice of values for the external momenta has no impact on the results for the quartic couplings. In the case of finite bare Green's functions, the $\MSbar$-renormalized Green's functions coincide with the bare ones in $DR$: 
\bea
\langle  A_{+ \, f_1}^{\dagger \alpha_1}(q_1)  A_{+ \, f_2}^{\dagger \alpha_2}(q_2) \, A_{+ \, f_3}^{\alpha_3}(q_3) A_{+ \, f_4}^{\alpha_4}(q_4)\rangle^{\MSbar, {\rm{1loop}}} - \langle  A_{+ \, f_1}^{\dagger \alpha_1}(q_1)  A_{+ \, f_2}^{\dagger \alpha_2}(q_2) \, A_{+ \, f_3}^{\alpha_3}(q_3) A_{+ \, f_4}^{\alpha_4}(q_4) \rangle^{DR, {\rm{1loop}}} = && \label{GFcontpppp}\nonumber \\
\langle  A_{- \, f_1}^{\dagger \alpha_1}(q_1)  A_{- \, f_2}^{\dagger \alpha_2}(q_2) \, A_{- \, f_3}^{\alpha_3}(q_3) A_{- \, f_4}^{\alpha_4}(q_4) \rangle^{\MSbar, {\rm{1loop}}} - \langle  A_{- \, f_1}^{\dagger \alpha_1}(q_1)  A_{- \, f_2}^{\dagger \alpha_2}(q_2) \, A_{- \, f_3}^{\alpha_3}(q_3) A_{- \, f_4}^{\alpha_4}(q_4) \rangle^{DR, {\rm{1loop}}} \phantom{=\,}  && \label{GFcontmmmm}\nonumber \\
&&\hspace{-8.25cm} = \frac{g^4}{64 \, \pi^2 \, N_c^2 } \, \frac{1}{\epsilon} \bigg (2 + 4 \, N_c^2 + \alpha \, (-2 \alpha + 3 \, (1+\alpha) \, N_c^2)  - 2 \, N_c \, N_f \bigg) \times 
  \nonumber \\
&&\hspace{-6.25cm} \bigg [ \delta_{f_1 f_3} \delta_{f_2 f_4}(-\delta^{\alpha_1 \alpha_3} \delta^{\alpha_2 \alpha_4}+N_c \, \delta^{\alpha_1 \alpha_4} \delta^{\alpha_2 \alpha_3}) \nonumber \\
&&\hspace{-6.25cm} + \delta_{f_1 f_4} \delta_{f_2 f_3}(-\delta^{\alpha_1 \alpha_4} \delta^{\alpha_2 \alpha_3}+N_c \, \delta^{\alpha_1 \alpha_3} \delta^{\alpha_2 \alpha_4}) \bigg ] \\
\langle  A_{+ \, f_1}^{\dagger \alpha_1}(q_1)  A_{+ \, f_2}^{\alpha_2}(q_2) \, A_{- \, f_3}^{\dagger \alpha_3}(q_3) A_{- \, f_4}^{\alpha_4}(q_4) \rangle^{\MSbar, {\rm{1loop}}} - \langle  A_{+ \, f_1}^{\dagger \alpha_1}(q_1)  A_{+ \, f_2}^{\alpha_2}(q_2) \, A_{- \, f_3}^{\dagger \alpha_3}(q_3) A_{- \, f_4}^{\alpha_4}(q_4) \rangle^{DR, {\rm{1loop}}}  && \label{GFcontppmm}\nonumber \\
&&\hspace{-8.25cm} = \frac{g^4}{64 \, \pi^2 \, N_c^2 } \, \frac{1}{\epsilon} \bigg( -2 - 4 \, N_c^2 + \alpha \,  (4 - N_c^2 + \alpha \, (-2 + 3 \, N_c^2)) + 2 \, N_c \, N_f \bigg) \times 
  \nonumber \\
&&\hspace{-6.25cm} \delta_{f_1 f_2} \delta_{f_4 f_3} ( N_c \, \delta^{\alpha_1 \alpha_3} \delta^{\alpha_4 \alpha_2} - \delta^{\alpha_1 \alpha_2} \delta^{\alpha_4 \alpha_3} )
\eea

The remaining Green's functions exhibit no pole parts. The difference between the renormalized Green's functions and the corresponding Green's functions regularized on the lattice allows us to deduce the one-loop renormalized quartic couplings. We emphasize that our results are mass-independent, as expected given that the difference between lattice and continuum Green's functions must be a polynomial of zeroth order in the momenta, and therefore, the mass could not be present for dimensional reasons.

For the sake of completeness we present the definition of the renormalization factor of the gauge coupling here:
\be
g \equiv g^B = Z_g^{-1}\,\mu^{\epsilon}\,g^R \label{g}, 
\ee
where $B$ stands for bare and $R$ for renormalized quantities and $\mu$ is the arbitrary scale appearing in DR with dimensions of inverse length. For one-loop calculations, the distinction between $g^R$ and $g^B$ is inessential in many cases; we will simply use $g$ in those cases. 

Note also that the components of the squark fields may mix at the quantum level, via a $2\times2$ mixing matrix ($Z_A$). We define the renormalization mixing matrix for the squark fields as follows:
\be
\label{condS}
\left( {\begin{array}{c} A^R_+ \\ A^{R\,\dagger}_- \end{array} } \right)= \left(Z_A^{1/2}\right)\left( {\begin{array}{c} A^B_+ \\ A^{B\,\dagger}_- \end{array} } \right).
\ee
However, in Ref.~\cite{Costa:2017rht} we found that in the $DR$ and $\MSbar$ scheme this $2\times2$ mixing matrix is diagonal. On the lattice this matrix is non diagonal; rather, it is found to be a real and symmetric matrix and the component $A_+ (A_-)$ mixes with $A_-^\dagger (A_+^\dagger)$. Thus, on the lattice the renormalization conditions are more complicated. 

The renormalization factor of the gauge parameter $Z_\alpha$ is defined as follows:
\be
\alpha^R = Z_\alpha^{-1} \, Z_u \, \alpha^B,
\label{alphaR}
\ee
where $Z_u$ is the renormalization factor of the gluon field, defined as: 
\be
u^R_\mu = \sqrt{Z_u} u_\mu^B.
\ee
By calculating the gluon self energy, it is found to be transverse, reflecting the gauge invariance of the theory. Since there is no longitudinal part for the gluon self energy, $Z_\alpha$  receives no one-loop contribution. 

In $DR$, we are interested in eliminating the pole parts in bare continuum Green's functions; this requires not only the renormalization factors of the fields, of the gauge coupling, $Z_g$, and of the gauge parameter, $Z_\alpha$, but requires a special treatment of the bare quartic coupling multiplying also with $Z_{\la_1}$. The quartic coupling is renormalized as follows: 
\bea
\lambda_1 = Z_{\lambda_1}^{-1} Z_g^{-2} \mu^{2 \epsilon} \, (g^R)^2 .
\label{gl}
\eea
At the lowest perturbative order, it holds that $Z^2_g Z_{\lambda_1} = 1$, and consequently, the renormalized quartic coupling aligns with the gauge coupling. 

Considering the example of the Green's function in $DR$ with four external squark fields $A_+$ and $A_+^\dagger$, the renormalization condition is expressed as follows:
\begin{align}
\langle   A_+(q_1) A^{\dagger}_+(q_2) A_+(q_3) A^{\dagger}_+(q_4) \rangle  \Big \vert^\MSbar =  (Z_A^{-2})_{++}   \nonumber \langle   A_+(q_1) A^{\dagger}_+(q_2) A_+(q_3) A^{\dagger}_+(q_4) \rangle\Big \vert^{\rm{bare}} \\
\label{renormL}
\end{align}
All appearances of coupling constants, and the gauge parameter in the right-hand side of Eq.~(\ref{renormL}) must be expressed in terms of their renormalized values, via Eq.~(\ref{g}), Eq.~(\ref{gl}), and Eq.~(\ref{alphaR}).

The renormalization factors in $DR$ which appear in the right-hand side of the renormalization condition are:
\bea
Z_{A_\pm}^{DR,\MSbar} &=& 1 + \frac{g^2\,C_F}{16\,\pi^2} \frac{1}{\epsilon}\left(-1 + \alpha \right)\\
Z_{g}^{DR,\MSbar}&=&1 + \frac{g^2\,}{16\,\pi^2} \frac{1}{\epsilon} \left(\frac{3}{2} N_c - \frac{1}{2}N_f \right)\\
Z_{u}^{DR,\MSbar}&=&1 + \frac{g^2\,}{16\,\pi^2} \frac{1}{\epsilon} \left[ \left( \frac{\alpha}{2} - \frac{3}{2} \right) N_c + N_f \right] \\
Z_{\alpha}^{DR,\MSbar}&=& 1 + {\cal O}(g^4)
\eea
where $C_F=(N_c^2-1)/(2\,N_c)$ is the quadratic Casimir operator in the fundamental representation.

Utilizing Eq.~(\ref{renormL}) for all bare Green's functions that have pole parts (see Eqs.~(\ref{GFcontpppp}) - (\ref{GFcontppmm})), we obtain the same value for the renormalization factor of $\la_1^{DR, \MSbar}$:
\bea
{Z_{\la_1}}^{DR,\MSbar} &=& 1 + {\cal O}(g^4)  
\label{ZlaDR}
\eea
As expected, we obtain the same value for the renormalization factor of $\lambda_1^{\text{DR}, \overline{\text{MS}}}$ by setting to zero the momenta of the squark fields that lie in the antifundamental representation, instead of those in the fundamental representation.

Eq.~(\ref{ZlaDR}) implies that, at the quantum level, the renormalization of the quartic coupling in DR remains unaffected by one-loop corrections. This observation carries significant implications for our comprehension of the renormalization scheme in SQCD. Furthermore, it indicates that the corresponding renormalization on the lattice will be finite. While terms proportional to $\lambda_2$ - $\lambda_5$ and $\lambda^F_1$ - $\lambda^F_5$ do not manifest themselves in the $\MSbar$ renormalization using DR, a finite mixture of these terms may emerge in $\MSbar$ on the lattice. We anticipate that the $\MSbar$ renormalization factors of gauge-invariant quantities will also be gauge-independent on the lattice, mirroring the behavior of $Z_{\lambda_1}^{\text{DR}, \overline{\text{MS}}}$.

\bigskip
Shifting our focus to $LR$, it is crucial to note that despite the diagonal nature of the renormalization matrix of the squark fields in the $\overline{\text{MS}}$ scheme and in $DR$, such simplicity does not carry over to the lattice. On the lattice, the mixing between squark components arises through the matrix $Z_A$, where the non-diagonal matrix elements are nonzero. Therefore, the renormalization conditions are not as straightforward as depicted in Eq.~(\ref{renormL}).

Taking into account the additional vertices on the lattice, we need to include additional one-loop Feynman diagrams to accurately calculate the fine-tuning of the quartic couplings. These diagrams are illustrated in Fig.~\ref{quarticFD}.

Now, on the lattice the renormalization condition up to $g^2$ will be given by: 
\begin{align}
\langle   A_+(q_1) A^{\dagger}_+(q_2) A_+(q_3) A^{\dagger}_+(q_4) \rangle  \Big \vert^\MSbar =&  
\langle \big((Z_A^{-1/2})_{++} A_+(q_1) + (Z_A^{-1/2})_{+-} A^{\dagger}_-(q_1) \big) \times \nonumber \\
&\big((Z_A^{-1/2})_{++} A^{\dagger}_+(q_2) + (Z_A^{-1/2})_{+-} A_-(q_2) \big) \times  \nonumber \\
&\big((Z_A^{-1/2})_{++} A_+(q_3) + (Z_A^{-1/2})_{+-} A^{\dagger}_-(q_3) \big) \times \nonumber \\ &\big((Z_A^{-1/2})_{++} A^{\dagger}_+(q_4) + (Z_A^{-1/2})_{+-} A_-(q_4) \big) \rangle\Big \vert^{\rm{bare}} 
\label{renormLlatt}
\end{align}
Analogous equations hold for the other Green's functions which involve the other matrix elements.

To provide a comprehensive overview, we revisit a collection of lattice results discussed in Ref.~\cite{Costa:2017rht}. 
\bea
\label{zpl}
Z_A^{LR,\MSbar} &=&  \openone - \,\frac{g^2\,C_F}{16\,\pi^2}\Bigg\{\Bigg[16.9216 - 3.7920\alpha-(1-\alpha)\log\left(a^2\,\bar\mu^2\right)\Bigg] \begin{pmatrix} 1 & 0\\ 0 & 1 \end{pmatrix} - 0.1623 \begin{pmatrix} 0 & 1\\ 1 & 0 \end{pmatrix}\Bigg\}\\
Z_g^{LR,\MSbar} &=&  1 + \gtilde\,\Bigg[ -9.8696 \frac{1}{N_c} + N_c \left( 12.8904  - \frac{3}{2} \log\left(a^2\,\bar{\mu}^2\right)\right)-\,N_f\left( 0.4811 - \frac{1}{2} \log(a^2\,\bar{\mu}^2)\right)\Bigg]
\label{zgl} \\
Z_{u}^{LR,\MSbar} &=&  1 + \frac{g^2\,}{16\,\pi^2} \left[19.7392\frac{1}{N_c}- N_c\left(18.5638 - 1.3863  \alpha + \left( -\frac{3}{2}+\frac{\alpha}{2}\right) \log\left(a^2\,\bar\mu^2\right)\right)+ N_f\left(0.9622   - \log\left(a^2\,\bar\mu^2\right)\right)\right] \nonumber \\
\\
Z_\alpha^{LR,\MSbar} &=& 1 + {\cal O}(g^4)
\eea

In the context of lattice Green's functions involving external fields as depicted in the action, certain terms exhibit negative powers of the lattice spacing. However, through the summation of specific diagram groups, namely diagrams (16, 20, 23, 26, 27), diagrams (18, 25), diagrams (17, 24) and diagrams (19, 21), these negative powers are canceled out, as required by gauge invariance. Furthermore, the lattice preserves only hypercubic invariance as a subset of Lorentz symmetry. Lorentz non-invariant terms of the external momentum emerge in the Green's functions, such as $\sum_\rho (q_{1 \rho})^4/(q_1^2)^2$ ; however, these terms also cancel out upon summation of all diagrams.
\\\\
Having checked that alternative choices of the external momenta give the same results for the differences between lattice and continuum Green's functions, we present them only for the case of two zero squark momenta. It is worth mentioning that the errors on our lattice expressions are smaller than the last shown digit.

\bea
\langle   A_{+ \, f_1}^{\dagger \alpha_1}(q_1)  A_{+ \, f_2}^{\dagger \alpha_2}(q_2) \, A_{+ \, f_3}^{\alpha_3}(q_3) A_{+ \, f_4}^{\alpha_4}(q_4) \rangle^{\MSbar, {\rm{1loop}}}  - \langle   A_{+ \, f_1}^{\dagger \alpha_1}(q_1)  A_{+ \, f_2}^{\dagger \alpha_2}(q_2) \, A_{+ \, f_3}^{\alpha_3}(q_3) A_{+ \, f_4}^{\alpha_4}(q_4)  \rangle^{LR, {\rm{1loop}}}  &=& \nonumber
\\
&&\hspace{-14.75cm} \frac{g^4}{16 \, \pi^2  } \,\frac{1}{N_c^2} \, \bigg[ 14.2135\, (1-N_c^2) \Big(\delta_{f_1 f_4} \delta_{f_2 f_3} \delta^{\alpha_1 \alpha_4} \delta^{\alpha_2 \alpha_3} + \delta_{f_1 f_3} \delta_{f_2 f_4} \delta^{\alpha_1 \alpha_3} \delta^{\alpha_2 \alpha_4} \Big)
  \nonumber \\
&&\hspace{-13.0cm}
+ \Big(30.3230 + 6.5648 \, \alpha - 1.8960 \, \alpha^2 - 8.0969 \, N_c^2 - 2.9562 \, \alpha \, N_c^2 + 2.5892 \, \alpha^2 \, N_c^2 - 1.8960 \, N_c \, N_f
  \nonumber \\
&&\hspace{-12.3cm} - \, (1/2  - \alpha^2 /2 + N_c^2 + 3\,\alpha \, N_c^2 /4 + 3\, \alpha^2 \, N_c^2/4 - N_c \, N_f/2) \, \log\left(a^2 \bar \mu^2 \right) \Big) \times   \nonumber \\
&&\hspace{-12.3cm}\Big(\delta_{f_1 f_4} \delta_{f_2 f_3} \, (-\delta^{\alpha_1 \alpha_4} \delta^{\alpha_2 \alpha_3}+N_c\,\delta^{\alpha_1 \alpha_3} \delta^{\alpha_2 \alpha_4}) + \delta_{f_1 f_3} \delta_{f_2 f_4} \, (-\delta^{\alpha_1 \alpha_3} \delta^{\alpha_2 \alpha_4}+N_c\,\delta^{\alpha_1 \alpha_4} \delta^{\alpha_2 \alpha_3}) \Big)\bigg] \\[0.7cm]
\langle   A_{+ \, f_1}^{\dagger \alpha_1}(q_1)  A_{+ \, f_2}^{\alpha_2}(q_2) \, A_{- \, f_3}^{\dagger \alpha_3}(q_3) A_{- \, f_4}^{\alpha_4}(q_4) \rangle^{\MSbar, {\rm{1loop}}}  - \langle  A_{+ \, f_1}^{\dagger \alpha_1}(q_1)  A_{+ \, f_2}^{\alpha_2}(q_2) \, A_{- \, f_3}^{\dagger \alpha_3}(q_3) A_{- \, f_4}^{\alpha_4}(q_4)  \rangle^{LR, {\rm{1loop}}}  &=& \nonumber
\\
&&\hspace{-14.75cm} \frac{g^4}{16 \, \pi^2  } \frac{1}{N_c^2} \bigg[-1.8246 \, \delta_{f_1 f_3} \delta_{f_4 f_2} \, \Big( (2 + N_c^2) \,  \delta^{\alpha_1 \alpha_3} \delta^{\alpha_4 \alpha_2} + N_c \, (-4 + N_c^2) \, \delta^{\alpha_1 \alpha_2} \delta^{\alpha_4 \alpha_3} \Big) 
  \nonumber \\
&&\hspace{-13.0cm}  + 14.2135 \, (1-N_c^2) \,
\delta_{f_1 f_2} \delta_{f_4 f_3} \, \delta^{\alpha_1 \alpha_2} \delta^{\alpha_4 \alpha_3}  \nonumber \\
&&\hspace{-13.0cm} -\Big(-26.5310 - 10.3568 \, \alpha + 1.8960 \, \alpha^2 + 6.1166 \, N_c^2 + 6.7483 \, \alpha \, N_c^2 - 2.5892 \, \alpha^2 \, N_c^2 - 1.8960 \, N_c \, N_f
  \nonumber \\
&&\hspace{-12.3cm} + \, (- 1/2  +  \alpha  -  \alpha^2/2  -  N_c^2 - \alpha \, N_c^2/4 + 3\,\alpha^2 \, N_c^2/4 + N_c \, N_f/2) \, \log\left(a^2 \bar \mu^2 \right) \Big) \times\nonumber \\
&& \hspace{-11.8cm} \delta_{f_1 f_2} \delta_{f_4 f_3} \, (-\delta^{\alpha_1 \alpha_2} \delta^{\alpha_4 \alpha_3} + N_c \,\delta^{\alpha_1 \alpha_3} \delta^{\alpha_4 \alpha_2}) \, \bigg]  \\[0.7cm]
\langle   A_{+ \, f_1}^{\alpha_1}(q_1) A_{- \, f_2}^{\alpha_2}(q_2) A_{+ \, f_3}^{\alpha_3}(q_3) A_{- \, f_4}^{\alpha_4}(q_4) \rangle^{\MSbar, {\rm{1loop}}}  - \langle    A_{+ \, f_1}^{\alpha_1}(q_1) A_{- \, f_2}^{\alpha_2}(q_2) A_{+ \, f_3}^{\alpha_3}(q_3) A_{- \, f_4}^{\alpha_4}(q_4)  \rangle^{LR, {\rm{1loop}}}  &=& \nonumber\\
&&\hspace{-14.25cm} \frac{g^4}{16 \, \pi^2  } \, \frac{1.8975}{N_c^2} \bigg( (2 + N_c^2) \times (\delta_{f_4 f_1} \delta_{f_2 f_3} \delta^{\alpha_4 \alpha_1} \delta^{\alpha_2 \alpha_3} + \delta_{f_4 f_3} \delta_{f_2 f_1} \delta^{\alpha_4 \alpha_3} \delta^{\alpha_2 \alpha_1}) + 
  \nonumber \\
&&\hspace{-12.65cm} N_c \, (-4 + N_c^2) \times (\delta_{f_4 f_1} \delta_{f_2 f_3} \delta^{\alpha_4 \alpha_3} \delta^{\alpha_2 \alpha_1} + \delta_{f_4 f_3} \delta_{f_2 f_1} \delta^{\alpha_4 \alpha_1} \delta^{\alpha_2 \alpha_3} ) \bigg) \\[0.7cm]
\langle   A_{+ \, f_1}^{\alpha_1}(q_1) A^{\dagger\alpha_2}_{+ \, f_2}(q_2) A_{+ \, f_3}^{\alpha_3}(q_3) A_{- \, f_4}^{\alpha_4}(q_4) \rangle^{\MSbar, {\rm{1loop}}}  - \langle   A_{+ \, f_1}^{\alpha_1}(q_1) A^{\dagger\alpha_2}_{+ \, f_2}(q_2) A_{+ \, f_3}^{\alpha_3}(q_3) A_{- \, f_4}^{\alpha_4}(q_4) \rangle^{LR, {\rm{1loop}}}  &=& \nonumber
\\
&&\hspace{-14.25cm} \frac{g^4}{16 \, \pi^2 } \,\frac{0.4913}{N_c^2} \bigg( -(2 + 0.1739 \, \alpha +  N_c^2 + 2.3923 \, \alpha \, N_c^2) \times (\delta_{f_4 f_1} \delta_{f_2 f_3} \delta^{\alpha_4 \alpha_1} \delta^{\alpha_2 \alpha_3} + \delta_{f_4 f_3} \delta_{f_2 f_1} \delta^{\alpha_4 \alpha_3} \delta^{\alpha_2 \alpha_1}) + 
  \nonumber \\
&&\hspace{-12.05cm} N_c \, (4 + 0.1739 \, \alpha -  N_c^2 + 2.3923 \, \alpha \, N_c^2) \times (\delta_{f_4 f_1} \delta_{f_2 f_3} \delta^{\alpha_4 \alpha_3} \delta^{\alpha_2 \alpha_1} + \delta_{f_4 f_3} \delta_{f_2 f_1} \delta^{\alpha_4 \alpha_1} \delta^{\alpha_2 \alpha_3} ) \bigg) 
\eea 
By combining the lattice expressions with the $\MSbar$-renormalized Green's functions calculated in the continuum (see Eq.~(\ref{renormLlatt})), we obtain the following renormalization factors and the coefficients of the counterterms: 
\bea
{Z_{\la_1}}^{LR,\MSbar} &=& 1 + \frac{g^2}{16 \, \pi^2} \bigg( \frac{35.0365}{N_c} - 25.0526 \, N_c - 2.8298 \, N_f \bigg) \\
{\la_2}^{LR,\MSbar} &=&  \frac{ g^4}{16 \, \pi^2} \bigg[ -0.9488 \, \bigg( 1 + \frac{2}{N_c^2} \bigg) \bigg] \\
{\la_3}^{LR,\MSbar} &=& \frac{ g^4}{16 \, \pi^2}  \bigg(  \frac{28.427}{N_c^2 }\bigg) \\
{\la_4}^{LR,\MSbar} &=& \frac{g^4}{16 \, \pi^2} \bigg( 16.0381 + \frac{3.6493}{N_c^2} \bigg)\\
{\la_5}^{LR,\MSbar} &=& \frac{ g^4}{16 \, \pi^2} \bigg[ 0.4913 \, \bigg(1 + \frac{2}{N_c^2} \bigg) \bigg] \\
{\la^F_1}^{LR,\MSbar} &=& \frac{ g^4}{16 \, \pi^2} \bigg( 28.427  \bigg) \\
{\la^F_2}^{LR,\MSbar} &=& \frac{g^4}{16 \, \pi^2} \bigg[0.9488 \, \bigg( \frac{4}{N_c}  -  N_c \bigg) \bigg]\\
{\la^F_3}^{LR,\MSbar} &=& \frac{g^4}{16 \, \pi^2} \bigg( -\frac{21.5121}{N_c} + 1.8246 \, N_c \bigg) \\
{\la^F_4}^{LR,\MSbar} &=& \frac{g^4}{16 \, \pi^2} \bigg[ 14.213 \, \bigg( -\frac{3}{N_c} +  N_c \bigg) \bigg] \\
{\la^F_5}^{LR,\MSbar} &=& \frac{g^4}{16 \, \pi^2} \bigg[0.4913 \, \bigg( -\frac{4}{N_c} +  N_c \bigg) \bigg]
\eea
We note that the above factors are gauge independent, as they should be in the $\MSbar$ scheme. Moreover, as anticipated from the continuum computations, these couplings acquire only finite (albeit nonzero) values.

We also present results for the case where the number of flavors is $N_f=1$. In this context, we provide the values for the five quartic terms, as there is no ``Fierz'' version with only one flavor. These terms are denoted as follows:
\bea
{Z_{\lambda_1}}^{LR,\overline{\text{MS}}}\Big{|}_{N_f=1} &=& 1 + \frac{g^2}{16 \, \pi^2} \bigg( -31.2567 + \frac{35.0366}{N_c} - 25.0526 \, N_c \bigg) \\
{\lambda_2}^{LR,\overline{\text{MS}}}\Big{|}_{N_f=1} &=&  \frac{g^4}{16 \, \pi^2} \bigg[-0.9488 \,\bigg( 1 + \frac{2}{N_c^2} - \frac{4}{N_c} + N_c \bigg)\ \bigg] \\
{\lambda_3}^{LR,\overline{\text{MS}}}\Big{|}_{N_f=1} &=& \frac{g^4}{16 \, \pi^2} \bigg(\frac{28.427}{N_c^2} - \frac{21.5121}{N_c} + 1.8246 \, N_c \bigg) \\
{\lambda_4}^{LR,\overline{\text{MS}}}\Big{|}_{N_f=1} &=& \frac{g^4}{16 \, \pi^2} \bigg[14.213 \bigg(1 + N_c -\frac{3}{N_c}\bigg) + 1.8246 \bigg(1 + \frac{2}{N_c^2}\bigg)\bigg] \\
{\lambda_5}^{LR,\overline{\text{MS}}}\Big{|}_{N_f=1} &=& \frac{g^4}{16 \, \pi^2} \bigg[0.4913 \, \bigg( 1 + \frac{2}{N_c^2} - \frac{4}{N_c} +  N_c \bigg) \bigg]
\eea
These results are essential for understanding the behavior of the theory in this particular flavor sector.

\newpage
\section{Outlook}
\label{summary}

In this work we calculate Green's functions with four external squark fields in SQCD, in the Wess-Zumino gauge. The fine-tunings of the quartic couplings are calculated employing Wilson fermions and gluons by studying the vertex corrections for the corresponding interactions. To extract these quantities in the $\MSbar$ scheme, we compute the relevant Green's functions in two regularizations: dimensional and lattice. The lattice calculations are the crux of this work; and the continuum calculations serve as a necessary introductory part, allowing us to relate our lattice results to the $\MSbar$ scheme.

The perturbative renormalization of these couplings signifies also the completion of all renormalizations (fields, masses, couplings) in the Wilson formulation~\cite{Costa:2017rht, Costa:2018mvb, Costa:2023cqv}. The results of this work will be particularly relevant for the setup and the calibration of lattice numerical simulations of SQCD. In the coming years, it is expected that simulations of supersymmetric theories will become ever more feasible and precise.

A natural extension of this work would be the perturbative calulations of all fine-tunings in SQCD and on the lattice using chirally invariant actions.  In particular, the overlap action can be used for gluino and quark fields, in order to ensure correct chiral properties. Simulating overlap fermions is a clear challenge, given the requirements in CPU time. On the other hand, the number of parameters which need fine-tuning is minimized, and this is a notable advantage for these kinds of calculations. Nevertheless, fixing the correct values of this minimal set of parameters still entails calculating a plethora of Green’s functions. 

\begin{acknowledgments}
 The project (EXCELLENCE/0421/0025) is implemented under the programme of social cohesion ``THALIA 2021-2027'' co-funded by the European Union through the Research and Innovation Foundation of Cyprus (RIF). The results are generated within the FEDILA software (project: CONCEPT/0823/0052), which is also implemented under the same ``THALIA 2021-2027'' programme, co-funded by the European Union through RIF. M.C. also acknowledges partial support from the Cyprus University of Technology under the ``POST-DOCTORAL'' programme.
\end{acknowledgments}

\end{document}